\documentclass{PNAStwo}
\usepackage{graphicx}
\usepackage{rotate}
\usepackage{epsfig}
\usepackage{cite}
\usepackage{amssymb,amsfonts,amsmath}

\begin{document}

\conflictofinterest{No conflicts of interest.}

\track{This paper was submitted directly to the PNAS office.}

\footcomment{Author Contributions: D.H. wrote the paper.  W.Y. performed computer simulations and analyzed data.}

\title{The Future of Social Experimenting}

\author{Dirk Helbing\thanks{To whom correspondence should be addressed. E-mail: dhelbing@ethz.ch}\affil{1}{ETH Zurich, Chair of Sociology, in particular of Modeling and Simulation, Clausiusstr. 50, 8092 Zurich, Switzerland},
\and Wenjian Yu\affil{1}{}}
\contributor{Submitted to Proceedings of the National Academy of Sciences
of the United States of America}
\maketitle
\begin{article}

\begin{abstract}
Recent lab experiments by Traulsen et al. \cite{Traulsen} for the spatial prisoner's dilemma suggest that exploratory behavior of human subjects prevents cooperation through neighborhood interactions over experimentally accessible time spans. This
indicates that new theoretical and experimental
efforts are needed to explore
the mechanisms underlying a number of
famous puzzles in the social sciences.
\end{abstract}


\dropcap{C}ooperation is the essence that keeps societies together. It is the basis of solidarity and social order. When humans stop cooperating, this can imply a war of everybody against everybody.
Understanding why and under what conditions humans cooperate is, therefore, one of the grand challenges of science \cite{chall}, particularly in social dilemma situations (where collective cooperation would be beneficial, but individual free-riding is the most profitable strategy). 

When humans have social dilemma kind of interactions with randomly changing partners, a ``tragedy of the commons'' \cite{Hardin}, i.e.  massive free-riding is expected to occur. But how are humans then able to create public goods (such as a shared culture or a public infrastructure) and build up functioning social benefit systems despite of their self-interest? Under what conditions will they be able to fight global warming collectively? To answer related research questions, scientists have experimentally studied, among other factors, the influence of spatial and network interactions on the level of cooperation in various games (including non-dilemmatic ones) \cite{Selten,Cassar,Kirchkamp}. 

In their laboratory experiments, Traulsen et al. \cite{Traulsen} have now 
implemented Nowak's and May's prisoner's dilemma in two-dimensional space \cite{Nowak}, where the size of the two-dimensional spatial grid, the number of interaction partners and the payoff parameters were modified for experimental reasons. The prisoner's dilemma describes interactions between pairs of individuals, where free-riding is tempting, and cooperation is risky. Therefore, if individuals interact with different people each time (as in the case of {\it well-mixed} interactions), everybody is predicted to end up free-riding. If the world was really like this, social systems would not work.

However, in computer simulations of the spatial prisoner's dilemma \cite{Nowak}, cooperative (``altruistic'') behavior is able to survive through spatial clustering of similar strategies. This finding, which can dramatically change the outcome of the system, is also called ``network reciprocity'' \cite{netrec}. When individuals have {\it neighborhood} interactions and imitate better performing neighbors unconditionally, this can create correlations between the behaviors of neighboring individuals. Such spatio-temporal pattern formation facilitates a co-evolution of the behaviors and the spatial organization of individuals, generating configurations that can promote cooperative behavior. In fact, some long-standing puzzles in the social sciences find a natural solution, when spatial interactions are taken into account. This includes the higher-than-expected level of cooperation in social dilemma situations and the spreading of costly punishment \cite{pun}. The question is, whether these theoretical findings are also supported by experimental data?

The recent results by Traulsen et al. \cite{Traulsen} do not indicate a significant effect of spatial neighborhood interactions on the level of cooperation. This is, because their experimental subjects did not show an unconditional imitation of neighbors with a higher payoff, as it is assumed in many game-theoretical models. In fact, it is known that certain game-theoretical results are sensitive to details of the model such as the number of interaction partners, the inclusion of self-interactions or not, or significant levels of randomness \cite{Szabo} (see Fig. 1). Moreover, people have proposed a considerable number of different strategy update rules, which matter as well. Besides unconditional imitation, these include the best response rule \cite{best}, multi-stage strategies such as tit for tat \cite{Axelrod}, win-stay-lose-shift rules \cite{win} and aspiration-dependent rules \cite{aspir}, furthermore probabilistic rules such as the proportional imitation rule \cite{prop}, the Fermi rule \cite{Fermi}, and the unconditional imitation rule with a superimposed randomness (``noise''). In addition, there are voter \cite{voter} and opinion dynamics models \cite{opin} of various kinds, which assume social influence. According to these, individuals would imitate behavioral strategies, which are more frequent in their neighborhood. So, how do individuals \textit{really} update their behavioral strategies?

Traulsen et al. find that the probability to cooperate increases with the number of cooperative neighbors as expected from the Asch experiment \cite{Asch}. Moreover, the probability of strategy changes increases with the payoff difference in a way that can be approximated by the Fermi rule \cite{Fermi}. In the case of two behavioral strategies only, it corresponds to the well-known multi-nomial logit model of decision theory \cite{MNL}. However, there is a discontinuity in the data as the payoff difference turns from positive to negative values, which may be an effect of risk aversion \cite{Kahne}. To describe the time-dependent level of cooperation, it is sufficient to assume unconditional imitation with a certain probability and strategy mutations otherwise, where the  mutation rate is surprisingly large in the beginning and exponentially decaying over time. 

Understanding the origin of this ``noise'' would be important to control it experimentally and to reveal effects that would otherwise be hidden in the randomness of the data. Do people make mistakes or do they \textit{choose} to behave in a noisy way? As Fig. 2 of Ref. \cite{Traulsen} shows, rather than quickly destroying cooperation, randomness leads to {\it more} cooperation in the experiment of Traulsen et al.  
than the unconditional imitation rule predicts. This goes along with a significantly higher average payoff than for the unconditional imitation rule (see Fig. 1). In other words, the random component of the strategy update is \textit{profitable} for the experimental subjects. This suggests that randomness in social systems may play a \textit{functional} role. 

Given that Traulsen et al. do not find effects of spatial interactions, do we have to say good bye to network reciprocity in social systems, despite the nice explanations it is offering? Probably not. The empirically confirmed spreading of obesity, smoking, happiness, and cooperation in social networks \cite{Chris} indicates that effects of imitating neighbors (also friends or colleagues) are relevant, but probably over longer time periods than 25 interactions. In fact, according to formula [3] of Traulsen et al., cooperation could spread after about 40 iterations, when the mutation rate has decreased to low enough values. Such an effect should occur, when self-interactions are taken into account (see Fig. 1). To make it observable experimentally, however, one would have to reduce the necessary number of iterations  by varying the experimental conditions. 

The particular value of the work by Traulsen et al. is that it facilitates more realistic computer simulations and, thereby, also allows one to determine payoff values and other model parameters, which are expected to produce interesting effects after an experimentally accessible number of iterations. Due to non-linear feedback effects, experimental games can have qualitatively different outcomes, which are hard to predict without extensive computer simulations  scanning the parameter space. Such parameter dependencies  could, in fact, explain some of the apparent inconsistencies between empirical observations in different areas of the world \cite{Herrmann} (at least when framing effects such as the {\it expected} level of reciprocity and their impact on the effective payoffs \cite{netrec} are taken into account). The progress in the social sciences by understanding the parameter-dependence of system behaviors would be enormous. While the effort to determine them {\it experimentally} is prohibitive, one could still check computationally predicted, parameter-dependent outcomes by targeted experiments. Hence, the future of social experimenting lies in the \textit{combination} of computational and experimental approaches, where computer simulations optimize the experimental setting and experiments are used to verify, falsify or improve the underlying model assumptions.

\begin{figure}[htbp]
\begin{center}\vspace*{-12mm}
\includegraphics[width=8cm]{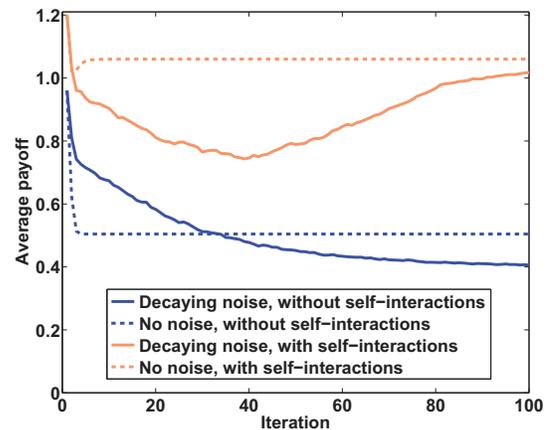}
\end{center}
\caption{Average payoff of all individuals in the spatial prisoner's dilemma with and without self-interactions, displayed over the number of iterations. It is clearly visible that the initial payoff drops quickly. In the noiseless case, the average payoff does not change anymore after a few iterations (see broken lines). This is, because the spatial configuration freezes. In contrast, in the case of decaying noise, the average payoff keeps changing (see solid lines). It is interesting that the average payoff, when no self-interactions are taken into account, is higher in the noisy case than in the noiseless one over the time period of the laboratory experiment by Traulsen et al., covering 25 iterations (see blue lines). The better performance in the presence of strategy mutations could be a possible reason for the high level of strategy mutations  observed by them. If self-interactions are considered (see orange lines), the average payoff recovers after about 40 iterations, which correlates with an increase in the level of cooperation (see Movie S1). To see this effect, experiments should be run over over at least 60 iterations, or the payoff parameters should be changed in such a way that the average payoff recovers earlier. It is conceivable, however, that experimental subjects would show a lower level of strategy mutations under conditions where noise does not pay off (in contrast to the experimental setting without self-interactions).}
\label{figure1}
\end{figure}

\end{article}
\end{document}